\documentclass[leqno,11pt,twoside]{article}
\usepackage[numbers]{natbib}
\usepackage{epsfig}

\usepackage{amsmath}
\usepackage{amssymb}
\usepackage[english]{babel}
\usepackage{multicol}

\def\ii{\int\limits_{\mathbb{R}}}
\def\jj{\int\limits_{-h_1}^{\eta}}
\def\jl{\int\limits_{-h_1}^{-l_1}}
\def\lj{\int\limits_{-l_1}^{\eta}}

\def\jjj{\int\limits_{\eta}^{h_2}}

\title{Hamiltonian Approach to Internal
Wave-Current Interactions in a Two-Media Fluid with a Rigid Lid}
\author{Alan Compelli, Rossen Ivanov}
\date{}

\begin{document}
\maketitle

\begin{abstract}
\noindent{\sc }
We examine a two-media 2-dimensional fluid system consisting of a lower medium bounded underneath by a flatbed and an upper medium with a free surface with wind generated surface waves but considered bounded above by a lid by an assumption that surface waves have negligible amplitude. An internal wave driven by gravity which propagates in the positive $x$-direction acts as a free common interface between the media. The current is such that it is zero at the flatbed but a negative constant, due to an assumption that surface winds blow in the negative $x$-direction, at the lid. We are concerned with the layers adjacent to the internal wave in which there exists a depth dependent current for which there is a greater underlying than overlying current. Both media are considered incompressible and having non-zero constant vorticities. The governing equations are written in canonical Hamiltonian form in terms of the variables, associated to the wave (in a presence of a constant current). The resultant equations of motion show that wave-current interaction is influenced only by the current profile in the 'strip' adjacent to the internal wave.

\end{abstract}



\section{Introduction}
\noindent{\sc }
Studies of internal waves, such as sharp temperature gradients called thermoclines which separate oceanic bodies of water which are at different temperatures, are of significant interest to climatologists, marine biologists, coastal engineers, etc.

The study of internal waves draws from previous single medium irrotational \cite{Zak}, \cite{BenjOlv}, \cite{Craig1}, \cite{Milder}, \cite{Miles0}, \cite{Miles} and rotational \cite{Constantin1}, \cite{Constantin2}, \cite{Constantin6}, \cite{Constantin7}, \cite{ConstantinEscher1}, \cite{TelesdaSilva}, \cite{NearlyHamiltonian}, \cite{Wahlen2} studies and from  appropriate studies of 2-media systems such as  \cite{CraigGuyKal}, \cite{CraigGuySul}, \cite{Compelli}, \cite{Compelli2}. However, these studies need to be extended to include the interaction between waves and currents.

Recent studies include the interaction between waves that propagate across the Pacific Ocean and the Equatorial Undercurrent (EUC) \cite{ConstRJo}, a Hamiltonian formulation describing the 2-dimensional nonlinear interaction between coupled surface waves, internal waves, and an underlying current with piecewise constant vorticity, in a two-layered fluid overlying a flat bed \cite{ConstIv2} and using shifted variables to transform a non-canonical wave-current system into a canonical system which has zero vorticity in the layers adjacent to the internal wave \cite{CompIv}. This study aims to provide a Hamiltonian formulation of a two-media bounded system which is rotational in the layers adjacent to the internal wave and hence show that wave-current interaction is influenced only by the current profile in this 'strip'.  
\section{Preliminaries}
\noindent{\sc }
The system under study consists of a 2-dimensional internal wave under the restorative action of gravity, which acts as a free common interface separating two fluid media, and a depth dependent current as per Figure 1. 
\begin{figure}
\centering
\includegraphics[width=0.8\textwidth]{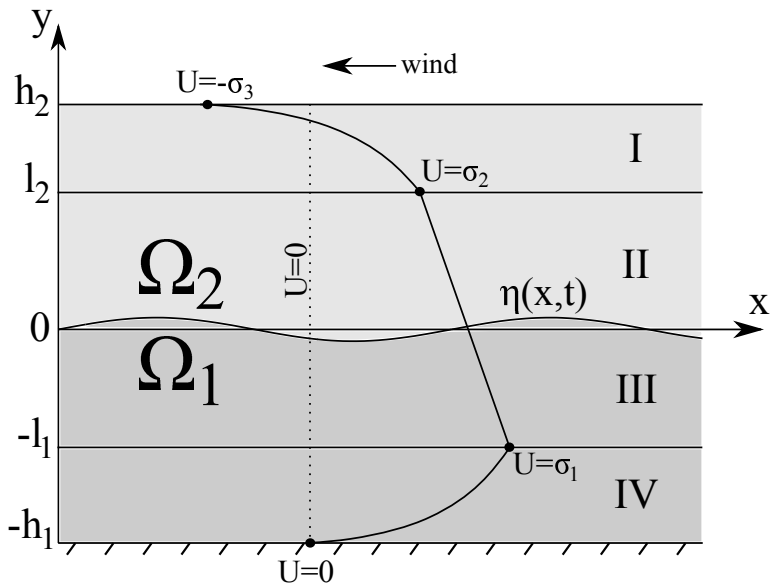}
\caption{System setup. The current profile in layers I and IV is arbitrary as we are only concerned with layers II and III as the internal wave is a free interface between these layers. Continuity of $U(y)$ is assumed in layers I and IV.}
\label{fig:1}       
\end{figure}

The medium underneath the internal wave is defined by the domain $\Omega_1=\{(x, y)\in\mathbb{R}^2: -h_1< y < \eta(x,t)\}$. This medium is bounded at the bottom by an impermeable flatbed at a depth $-h_1$. The medium above the internal wave is defined by the domain $\Omega_2=\{(x, y)\in\mathbb{R}^2: \eta(x,t)< y < h_2\}$. This medium is regarded as being bounded on top by an impermeable lid at a height $h_2$, but in reality is a free surface with negligible wave amplitude. Throughout the article the subscript $1$ will be used to mean evaluation for the lower medium $\Omega_1$, subscript $2$ means evaluation for the upper medium $\Omega_2$,  subscript $i=\{1,2\}$ means evaluation for both media and subscript $c$ will be used to denote evaluation at the common interface.
Non-lateral velocity flow is described by ${\bf{V}}_i(x,y,z)=(u_i,v_i,0)$. The arbitrary periodic function $\eta(x,t)$ describes the elevation of the internal wave, i.e. $y=\eta$ is the equation of the internal wave. We define the mean of $\eta$ to be the shear surface at $y=0$ with the centre of gravity in the negative $y$-direction. 

A depth dependent current $U_1(y)$ flows in $\Omega_1$ and, correspondingly, $U_2(y)$ flows in $\Omega_2$. Currents are described for the system under study via the  continuous function $U_i(y)$ as
\begin{equation}
\label{U_depth}
  U_i(y)=      \left\lbrace
        \begin{array}{lcl}
        -\sigma_3, \qquad y=h_2\mbox{ (lid)}
        \\
        \sigma_2, \qquad y=l_2
        \\
        \gamma y+\kappa,\qquad  l_2\ge y \ge -l_1 \mbox{ (layers II and III)}
        \\
        \sigma_1, \qquad y=-l_1
        \\
        0, \qquad   y = -h_1\mbox{ (flatbed)}
        \end{array}
        \right.
\end{equation}
for the positive constants $\sigma_1$, $\sigma_2$, $\sigma_3$, $\kappa$, $l_i$, $h_i$, $\gamma$ and $\kappa$, where $\kappa$ is the velocity of the time-independent current at $y=0$ and $\gamma$ is the non-zero constant vorticity for layers II and III, noting that the current is arbitrary in layers I and IV (however represented by a continuous function everywhere).

We consider a velocity field which is defined by:
\begin{equation}
\label{phi_def}
        \left\lbrace
        \begin{array}{lcl}
        u_i  =  \tilde{\varphi}_{i,x} +U_i(y)
        \\
        v_i =  {\tilde{\varphi}}_{i,y}.
        \end{array}
        \right.
\end{equation}
We have separated the wave and current contributions to the velocity and so we define ${\tilde{\varphi}}_i$ as the wave velocity potential for $\Omega_i$ and in particular the velocity components in layers II and III are [22]
\begin{equation}
\label{phitilda_def}
        \left\lbrace
        \begin{array}{lcl}
        u_i  =  \tilde{\varphi}_{i,x} +\gamma y+ \kappa
        \\
        v_i =  {\tilde{\varphi}}_{i,y}.
        \end{array}
        \right.
\end{equation}
Additionally, the stream function $\psi_i$ is introduced, defined by:
\begin{equation}
\label{psi_def}
        \left\lbrace
        \begin{array}{lcl}
        u_i  = \psi_{i,y}
        \\
        v_i =  -\psi_{i,x}.
        \end{array}
        \right.
\end{equation}

\noindent $\rho_1$ and $\rho_2$ are the respective constant densities of the lower and upper media and stability is given by the immiscibility condition
\begin{alignat}{2}
\label{stability}
\rho_1>\rho_2.
\end{alignat}

The rotationality of the layers II and III is given by the condition
\begin{alignat}{2}
\label{rotat}
\gamma<0\Leftrightarrow\sigma_1>\sigma_2
\end{alignat}
ensuring non-zero vorticity in this region. Alternatively $\sigma_2>\sigma_1$ could also be considered for $\gamma>0$.

We assume that for large $|x|$ the amplitude of $\eta$ attenuates and hence make the following assumptions
\begin{alignat}{2}
\label{Assump1}
\lim_{|x|\rightarrow \infty}\eta(x,t)=0,
\end{alignat}
\begin{alignat}{2}
\label{Assump1a}
\lim_{|x|\rightarrow \infty}{\tilde{\varphi}}_i(x,y,t)=0,
\end{alignat}
and
\begin{alignat}{2}
\label{Assump1b}
-l_1< \eta(x,t) < l_2\mbox{ for all $x$ and $t$},
\end{alignat}

\noindent i.e. the wave is localised in the {\it strip}.

We have the following equation (Euler's equation)
\begin{alignat}{2}
\label{Euler_1}
\nabla\Big(( \varphi_{i,t} )_c+\frac{1}{2}(\nabla \psi_i)_c^2-\gamma\psi_i\Big)=\nabla\Big(-\frac{p_i}{\rho_1} -g\eta\Big)
\end{alignat}
where $p_i$ is the dynamic pressure, $g$ is the acceleration due to gravity (where $y$ points in the opposite direction to the center of gravity) and $\nabla=(\partial_x,\partial_y)$. The following Bernoulli condition (cf. \cite{Compelli2}) at the interface follows from Euler's equation and assumptions (\ref{Assump1}) and (\ref{Assump1a}):
\begin{alignat}{2}
\label{Bernoulli0}
\rho_1\Big(( \varphi_{1,t} )_c+\frac{1}{2}(\nabla \psi_1)_c^2-\gamma\chi_1 +g\eta\Big)=\rho_2\Big(( \varphi_{2,t} )_c+\frac{1}{2}(\nabla \psi_2)_c^2-\gamma\chi_2 +g\eta\Big)+f(t)
\end{alignat}
where $\chi_i$ is the stream function evaluated at the interface. Since the two media do not mix, $\chi_1 = \chi_2 \equiv \chi$. Moreover,  $f(t)$ is an arbitrary function of time and depends on how the potentials are defined at $\pm\infty$. Clearly such a function can be absorbed in the definition of the wave potentials, but we will keep it separate for further convenience. We know by comparing (\ref{phitilda_def}) and (\ref{psi_def}) that
\begin{alignat}{2}
\frac{1}{2}(\nabla \psi_i)_c^2
=
\frac{1}{2}({\tilde{\varphi}}_{i,x})_c^2+\frac{1}{2}({\tilde{\varphi}}_{i,y})_c^2 + \frac{1}{2}(\gamma\eta+\kappa)^2+(\tilde{\varphi}_{i,x})_c \gamma \eta+\kappa (\tilde{\varphi}_{i,x})_c
\end{alignat}
and hence we can express the Bernoulli condition in terms of wave and current components only as
\begin{multline}
\label{BernoulliNew}
(\rho_1 {\tilde{\varphi}}_{1,t}-\rho_2 {\tilde{\varphi}}_{2,t} )_c +\kappa(\rho_1\tilde{\varphi}_{1,x}-\rho_2 \tilde{\varphi}_{2,x})_c 
+\frac{\rho_1}{2}|\nabla{\tilde{\varphi}}_{1}|_c^2
-\frac{\rho_2}{2}|\nabla{\tilde{\varphi}}_{2}|_c^2 \\
 +  \frac{1}{2}(\rho_1-\rho_2)(\gamma\eta+\kappa)^2
+\gamma\eta (\rho_1\tilde{\varphi}_{1,x} -\rho_2\tilde{\varphi}_{2,x})_c \\
-(\rho_1-\rho_2)\gamma\chi
 +(\rho_1-\rho_2)g\eta=f(t).
\end{multline}

The terms with $\gamma$ and $\kappa$ are due to the wave-current interaction. For example, the second term is due to overall translation leading to a shift $ \partial_t \rightarrow \partial_t + \kappa \partial_x $. The equation suggests the introduction of the variable \begin{alignat}{2}
\label{xi_maindef}
\xi:=\rho_1\xi_1-\rho_2\xi_2,
\end{alignat} 

\noindent where   
\begin{alignat}{2}
\xi_i:=(\tilde{\varphi}_i)_c=\tilde{\varphi}_i(x,\, \eta(x,\,t),\,t).
\end{alignat}
We also have the following  kinematic boundary conditions at the interface, using the velocity representations \eqref{phitilda_def}
\begin{equation}
\label{KBC}
        \left\lbrace
        \begin{array}{lcl}
        \eta_t  + \eta_x \big(\gamma \eta+(\tilde{\varphi}_{i,x})_c +\kappa\big) + (\tilde{\varphi}_{i,y})_c=0
        \\
        (\tilde{\varphi}_{1,y})_b=(\tilde{\varphi}_{2,y})_l=0
        \end{array}
        \right.
\end{equation}
noting that ${\bf{V_1}}(x,-h_1,0)=(u_1,0,0)$ and ${\bf{V_2}}(x,h_2,0)=(u_2,0,0)$, where the subscripts $b$ and $l$ denote evaluation at the bottom (lower boundary) and lid (upper boundary) respectively.
\section{Hamiltonian Formulation}
\noindent{\sc }
If we consider the system under study as an irrotational system the Hamiltonian, $H$, is given by the sum of the kinetic and potential energies as:
\begin{alignat}{2}
H= \frac{\rho_1}{2}\ii  \jj (u_1^2+v_1^2)dydx+ \frac{\rho_2}{2}\ii  \jjj  (u_2^2+v_2^2)dydx
+\frac{1}{2}(\rho_1-\rho_2)\ii g\eta^2 dx.
\end{alignat}
The kinetic energy term for $\Omega_1$ is
\begin{alignat}{2}
K_1= \frac{\rho_1}{2}\ii  \jj (u_1^2+v_1^2)dydx
\end{alignat}
which we can split into layers IV and III, respectively, as
\begin{alignat}{2}
K_1=\frac{\rho_1}{2}\ii  \jl (u_1^2+v_1^2)dydx+\frac{\rho_1}{2}\ii  \lj (u_1^2+v_1^2)dydx.
\end{alignat}
For layer IV the kinetic energy is
\begin{multline}
\frac{\rho_1}{2}\ii  \jl (u_1^2+v_1^2)dydx=
\frac{\rho_1}{2}\ii \jl ({\tilde{\varphi}}_{1,x})^2dydx
+\frac{\rho_1}{2}\ii  \jl ({\tilde{\varphi}}_{1,y})^2 dydx\\
+\frac{\rho_1}{2}\ii  \jl \gamma^2 y^2dydx
+ \frac{\rho_1}{2}\ii  \jl U_1^2dydx
+\rho_1\ii \jl \tilde{\varphi}_{1,x} \gamma ydydx\\
+\rho_1\ii \jl \gamma  U_1 ydydx
+\rho_1\ii \jl U_1 \tilde{\varphi}_{1,x}dydx.
\end{multline}
However, terms 3-7 combine to produce a constant which is irrelevant in terms of dynamic considerations (does not contribute to the variations with respect to the field variables). Moreover $\ii \eta(x',t) dx'=0$ (the mean deviation is by definition zero) and the fields vanish at $x=\pm \infty$ so that integration of total $x-$ derivatives produces zero, thus  
\begin{multline}
\frac{\rho_1}{2}\ii  \jl (u_1^2+v_1^2)dydx=
\frac{\rho_1}{2}\ii \jl ({\tilde{\varphi}}_{1,x})^2dydx
+\frac{\rho_1}{2}\ii  \jl ({\tilde{\varphi}}_{1,y})^2 dydx.
\end{multline}
For layer III the kinetic energy is
\begin{multline}
\frac{\rho_1}{2}\ii  \lj (u_1^2+v_1^2)dydx=
\frac{\rho_1}{2}\ii \lj ({\tilde{\varphi}}_{1,x})^2dydx
+\frac{\rho_1}{2}\ii  \lj ({\tilde{\varphi}}_{1,y})^2 dydx\\
+\frac{\rho_1}{2}\ii  \lj (\gamma y+\kappa)^2 dydx
+\rho_1\ii \lj \tilde{\varphi}_{1,x} \gamma ydydx
+\rho_1\ii \lj \kappa \tilde{\varphi}_{1,x}dydx.
\end{multline}
We write
\begin{alignat}{2}
\frac{\rho_1}{2}\ii  \lj (\gamma y+ \kappa )^2 dydx
=\frac{\rho_1}{6\gamma}\ii (\gamma \eta+\kappa)^3dx
\end{alignat}
noting that $\ii (\gamma \eta+\kappa)^3dx$ can be properly re-normalised as $\ii ((\gamma \eta+\kappa)^3-\kappa^3)dx$ as the variation in $\ii \kappa^3 dx$ is zero. 

We introduce the Dirichlet-Neumann operator $G_i(\eta)$ (see \cite{Craig1}, \cite{CraigGuySul}) given by
\begin{alignat}{2}
G_i(\eta)\xi_i=(\partial_{{{\bf{n}}_i}}\tilde{\varphi}_i)\sqrt{1+(\eta_x )^2},
\end{alignat}
where $\partial_{{{\bf{n}}_i}}\tilde{\varphi}_i$ is the normal derivative of the velocity potential $\tilde{\varphi}_i$, at the interface, for an outward normal ${{\bf{n}}_i}$, and also define \cite{CraigGuyKal}
\begin{alignat}{2}
\label{B_def}
B:=\rho_1 G_2(\eta)+\rho_2 G_1(\eta).
\end{alignat}
Thus we can determine that
\begin{alignat}{2}
\label{B_xi}
        \left\lbrace
        \begin{array}{lcl}
        \xi_1=B^{-1}\big(G_2(\eta)\xi\big)
        \\
        \xi_2=B^{-1}\big(-G_1(\eta)\xi\big)
        \end{array}
        \right.
\end{alignat}
The integral with $\rho_1 \kappa \tilde{\varphi}_{1,x}$ term, using the Leibniz integral rule with varying limits (cf. \cite{Compelli}), can be written as
\begin{alignat}{2}
\rho_1\ii  \lj \kappa\tilde{\varphi}_{1,x}dydx=-\rho_1\kappa\ii \xi_1 \eta_x dx
\end{alignat}
and the $\rho_1\gamma y \tilde{\varphi}_{1,x}$ term as
\begin{alignat}{2}
\label{intcalc_6}
\rho_1\ii  \jj\gamma y\tilde{\varphi}_{1,x}  dydx&=-\rho_1\ii  \gamma\xi_1\eta \eta_xdx
\end{alignat}
and hence we write the Hamiltonian for $\Omega_1$ as
\begin{multline}
H_1= 
\frac{\rho_1}{2}\ii \jj | \nabla{\tilde{\varphi}}_{1}|^2dydx
+\frac{\rho_1}{2}\ii g\eta^2 \, dx\\
+\frac{\rho_1}{6\gamma}\ii (\gamma \eta+\kappa)^3dx
-\rho_1\ii  \gamma\xi_1\eta \eta_xdx
-\rho_1\kappa\ii \xi_1 \eta_x dx.
\end{multline}
We follow the same procedure for $\Omega_2$ to obtain
the corresponding energy as
\begin{multline}
H_2= 
\frac{\rho_2}{2}\ii \jjj |\nabla {\tilde{\varphi}}_{2}|^2dydx
-\frac{\rho_2}{2}\ii   g\eta^2 dydx\\
-\frac{\rho_2}{6\gamma}\ii (\gamma \eta+\kappa)^3dx
+\rho_2\ii  \gamma\xi_2\eta \eta_xdx
+\rho_2\kappa\ii \xi_2 \eta_x dx.
\end{multline}
The total energy is therefore $H=H_1+H_2$ or in terms of ($\eta$, $\xi$)
\begin{multline}
\label{Ham_conj}
H(\eta,\xi)= \frac{1}{2}\ii  \xi \big(G_1(\eta) B^{-1}G_2(\eta)\big)\xi \,dx+\frac{\rho_1-\rho_2}{2}\ii g\eta^2 \, dx-\kappa\ii \xi \eta_x dx\\
-\ii  \gamma\eta\eta_x \xi dx
+\frac{(\rho_1-\rho_2)}{6\gamma}\ii (\gamma \eta+\kappa)^3dx.
\end{multline}
Defining the Hamiltonian which has no current or vorticity components, $H_0$, as
\begin{alignat}{2}
\label{Ham_conj}
H_0(\eta,\xi)= \frac{1}{2}\ii  \xi \big(G_1(\eta) B^{-1}G_2(\eta)\big)\xi \,dx+(\rho_1-\rho_2)\frac{1}{2}\ii g\eta^2 \, dx\end{alignat}
we can write
\begin{alignat}{2}
\label{Ham_conj_new_old}
H(\eta,\xi)= H_0-\kappa\ii \xi \eta_x dx-\ii  \gamma\eta\eta_x \xi dx
+\frac{(\rho_1-\rho_2)}{6\gamma}\ii (\gamma \eta+\kappa)^3dx.
\end{alignat}
The equations of motion can be written in Hamiltonian form as follows. From (\ref{KBC}) the dynamic boundary condition
\begin{alignat}{2} \label{34}
\eta_t  &=-  \gamma \eta\eta_x+(\tilde{\varphi}_{i,x})_c \eta_x-\kappa\eta_x - (\tilde{\varphi}_{i,y})_c\notag\\
&=\delta_{\xi} H_0-\kappa\eta_x-  \gamma \eta\eta_x= \delta_{\xi} H.
\end{alignat}

We note that the quantities in the Bernoulli condition (\ref{BernoulliNew}) are
\begin{alignat}{2}
\label{Bern1}
\rho_1({\tilde{\varphi}_{1,x}})_c-\rho_2({\tilde{\varphi}_{2,x}})_c&= \xi_x-(\rho_1{\tilde{\varphi}_{1,y}}-\rho_2 {\tilde{\varphi}_{2,y}})_c\eta_x\\
\label{Bern2}
\rho_1({\tilde{\varphi}_{1,t}})_c-\rho_2({\tilde{\varphi}_{2,t}})_c&= \xi_t-(\rho_1{\tilde{\varphi}_{1,y}}-\rho_2 {\tilde{\varphi}_{2,y}})_c\eta_t.
\end{alignat}
\noindent and we can write it as  

 \begin{multline}
\label{BernoulliNew2}
\xi_t - (\rho_1 {\tilde{\varphi}}_{1,y}-\rho_2 {\tilde{\varphi}}_{2,y} )_c(\eta_t+(\gamma \eta +\kappa) \eta_x)  
+\frac{\rho_1}{2}|\nabla{\tilde{\varphi}}_{1}|_c^2
-\frac{\rho_2}{2}|\nabla{\tilde{\varphi}}_{2}|_c^2 \\
+(\gamma \eta +\kappa)\xi_x  +  \frac{1}{2}(\rho_1-\rho_2)(\gamma\eta+\kappa)^2
-(\rho_1-\rho_2)\gamma\chi
 +(\rho_1-\rho_2)g\eta=f(t).
\end{multline}

\noindent or due to \eqref{34} as

 \begin{multline}
\label{BernoulliNew3}
\xi_t - (\rho_1 {\tilde{\varphi}}_{1,y}-\rho_2 {\tilde{\varphi}}_{2,y} )_c({\tilde{\varphi}}_{i,x}  \eta_x-{\tilde{\varphi}}_{i,y})_c  
+\frac{\rho_1}{2}|\nabla{\tilde{\varphi}}_{1}|_c^2
-\frac{\rho_2}{2}|\nabla{\tilde{\varphi}}_{2}|_c^2+(\rho_1-\rho_2)g\eta \\
+(\gamma \eta +\kappa)\xi_x +  \frac{1}{2}(\rho_1-\rho_2)(\gamma\eta+\kappa)^2
-(\rho_1-\rho_2)\gamma\chi
 =f(t).
\end{multline}

Noting the 'usual', not related to the current terms,  \begin{multline}
\label{BernoulliNew4}
\xi_t + \delta_{\eta} H_0 +(\gamma \eta +\kappa)\xi_x +  \frac{1}{2}(\rho_1-\rho_2)(\gamma\eta+\kappa)^2
-(\rho_1-\rho_2)\gamma\chi
 =f(t).
\end{multline}

\noindent and finally, from \eqref{Ham_conj_new_old}   \begin{equation}
\label{BernoulliNew5}
\xi_t + \delta_{\eta} H-(\rho_1-\rho_2)\gamma\chi =f(t).
\end{equation}

The equation for $\xi_t$ is given up to an arbitrary function of time because the Hamiltonian can always be 'renormalised' by a term $-f(t)\ii \eta dx$ which has a variation of $-f(t)$ with respect to $\eta$ but is otherwise zero by definition. Thus, for the renormalised Hamiltonian \begin{equation}
\label{BernoulliNew5}
\xi_t = - \delta_{\eta} H+(\rho_1-\rho_2)\gamma\chi .
\end{equation}

Since $\chi = - \int _{-\infty}^{x}\eta_t (x',t) dx'=-\int _{-\infty}^{x}\delta_{\xi}H dx', $ after a change of variables \cite{Wahlen2} via the transformation $(\eta,\xi)\rightarrow(\eta,\zeta)$  
\begin{alignat}{2}
\label{vartrans}
\xi\rightarrow\zeta=\xi-\frac{(\rho_1 - \rho_2)\gamma}{2} \int_{-\infty}^{x} \eta(x',t)\,dx'.
\end{alignat}

\noindent the system acquires a canonical Hamiltonian form:

\begin{equation}
\label{EOMsys}
        \left\lbrace
        \begin{array}{lcl}
        \eta_t=\delta_{\zeta} H
        \\
        \zeta_t=-\delta_{\eta} H
        \end{array}
        \right.
\end{equation} 

In conclusion we have shown that the wave-current is influenced only by the current profile in the 'strip' (layers II and III), i.e. outside this region the continuous current is arbitrary.

\section{Conclusion}
\noindent{\sc }

The governing equations of a system of two-media,  bounded on top by a lid and on the bottom by a flatbed, with an internal wave providing a free common interface and with a depth dependent current were written in  a canonical Hamiltonian form in terms of the 'wave'-related variables  $(\eta,\zeta).$ 

It was then shown that the wave-current interactions are influenced only by the current profile in the 'strip', and do not depend on the current profile in the other layers.


\section*{Acknowledgements}
\noindent{\sc }
A.C. is funded by the Fiosraigh Scholarship Programme of Dublin Institute of Technology.  The support of the FWF Project I544-N13 ``Lagrangian kinematics of water waves'' of the Austrian Science Fund is gratefully acknowledged by R.I. The authors are grateful to Prof. A. Constantin for many valuable discussions.

\begin{multicols}{2}

\noindent{\sc }\emph{Alan Compelli\\
School of Mathematical Sciences\\
Dublin Institute of Technology\\
Kevin Street, Dublin 8, Ireland}
\columnbreak
 
\noindent{\sc }\emph{Rossen Ivanov\\
School of Mathematical Sciences\\
Dublin Institute of Technology\\
Kevin Street, Dublin 8, Ireland\\
email: {\bf{rossen.ivanov@dit.ie}}\\
and\\
Faculty of Mathematics\\
Oskar-Morgenstern-Platz 1\\
 University of Vienna\\
1090 Vienna, Austria\\
}
\end{multicols}


\end{document}